\documentclass[12pt, prd, showpacs,showkeys,superscriptaddress,nofootinbib]{revtex4}
\usepackage{amssymb}
\usepackage{amsmath}
\usepackage{setspace}

\usepackage{color}
\definecolor{DarkGreen}{rgb}{0,0.4,0}
\definecolor{DeepBlue}{rgb}{0,0,.4}
\usepackage[naturalnames=true,colorlinks=true,
citecolor=DarkGreen,urlcolor=blue,linkcolor=DeepBlue]{hyperref}

\begin{document}

\title{Ba\~{n}ados-Silk-West effect with nongeodesic particles:\\
Nonextremal horizons}
\author{I. V. Tanatarov}
\email{igor.tanatarov@gmail.com}
\affiliation{Kharkov Institute of Physics and Technology,\\
1 Akademicheskaya, Kharkov 61108, Ukraine}
\affiliation{Department of Physics and Technology,\\
Kharkov V.N. Karazin National University,\\
4 Svoboda Square, Kharkov 61022, Ukraine}
\author{O. B. Zaslavskii}
\email{zaslav@ukr.net}
\affiliation{Department of Physics and Technology,\\
Kharkov V.N. Karazin National University,\\
4 Svoboda Square, Kharkov 61022, Ukraine}
\affiliation{Institute of Mathematics and Mechanics,\\
Kazan Federal University,\\
18 Kremlyovskaya St., Kazan 420008, Russia}

\begin{abstract}
If two particles collide near a black hole, the energy in their center of
mass can, under certain conditions, grow unbounded. This is Ba\~{n}ados-Silk-West effect. We show that this effect retains its validity even if some force acts on a particle, provided some reasonable and weak restrictions are imposed on this force. In the present paper we discuss the case of nonextremal horizons. The result under discussion is similar to that
for extremal horizons considered in our previous paper. The problem can be
viewed both in its own right and as a simple setup in which this force models in the first approximation the complicated gravitational self-force.
\end{abstract}

\maketitle


\newpage

\section{Introduction}

It was shown by Ba\~{n}ados, Silk and West (hereafter, BSW) \cite{ban} that
near-horizon collision of two particles moving towards the extremal Kerr
black hole can result in the indefinite growth of their energy $E_{c.m.}$ in
the center of mass frame. Soon after this observation, several arguments
were pushed forward against the possibility of physical realization of this
effect. The first one consisted in that realistic astrophysical black holes
cannot be exactly extremal \cite{ted}. However, it was refuted since the BSW
effect was extended to the nonextremal horizons of Kerr \cite{gp} and other
stationary axially symmetric black holes \cite{prd}. Another objection was
based on the role of gravitational radiation, which was assumed to bound the
BSW effect \cite{berti}. However, recent studies showed that under rather
general and weak assumptions, the BSW effect survives even if a force
(modeling the effect of radiation, backreaction, etc.) acts on the
particles \cite{rad}. This was obtained for the extremal horizons. Now our
goal is to consider the possibility of the BSW effect near horizons of
nonextremal black holes when particles move under the action of some force. 

It is worth mentioning that there are two kinds of potential limitations on
the BSW effect. The first one concerns the possibility to get unbounded $E_{c.m.}$, which involves only processes in the immediate vicinity of the horizon. The second kind is related to the issue of astrophysical relevance and potential observational significance of the BSW effect. In this regard the behavior of debris after collision in the asymptotically flat region is also important. The observable energy and mass at infinity for the extremal Kerr metric were found to be restricted by some upper limits in \cite{p} (built on \cite{pir3}), and a similar result was obtained in \cite{j}. Extension to more general ``dirty'' black holes was done in \cite{z}. A
separate question is whether fluxes at infinity can exceed the sensitivity of
modern devices. In general, the situation remains controversial -- \cite{mc}, 
\cite{com}. Here we discuss only the first kind of limitation, having
obvious theoretical value, and put aside the second kind, which is important
but needs separate further treatment. Up to date, there are already many
other different aspects of the BSW effect that remain beyond the scope of
the present paper. 

In this paper we discuss the BSW effect under the action of a force of a
rather generic character. When the corresponding results are applied to the
question regarding gravitational self-force, important reservations are in
order. The true gravitational self-force differs from simple external
force and depends on the particle's position, velocity, etc. in a highly
nonlinear way. The full analysis also needs to
consider the motion of the particle in nonstationary and nonaxisymmetric background \cite{bar}. In this sense, the present paper, as well as our previous one \cite{rad}, should be considered the first step to understanding BSW effect under the action of gravitational self-force,
which we model with the help of a``usual'' force.

\section{General formulas}

Let us consider the axially symmetric stationary black hole metric%
\begin{equation}
ds^{2}=-N^{2}dt^{2}+g_{\phi }(d\phi -\omega dt)^{2} +\frac{dr^{2}}{A}%
+g_{z}dz^{2}.
\end{equation}%
All metric coefficients do not depend on $t$ and $\phi $. The lapse function 
$N$ turns to zero at the horizon, where $N^2 \sim A$.

Let us consider particle's motion in the equatorial plane and, for
simplicity, put the mass of each particle equal to unity $m_{1,2}=1$.
Hereafter we will use two frames, which are convenient for description of
the processes near horizon -- the ``OO frame'', which is attached to an
observer orbiting a black hole with the zero angular momentum \cite{72} and
the ``FO frame'', which is attached to an observer falling into the black
hole.

It is convenient to parametrize a particle's for-velocity $u^\mu$ by its
energy $E$ and angular momentum (here $\mu=t,\phi,r,z$) 
\begin{equation}
u_\mu =(-E, L, u_r , 0).  \label{EL}
\end{equation}
Then the normalization condition for the four-velocity $u^\mu u_\mu =-1$ can
be presented as 
\begin{equation}
\frac{1}{A}(u^{r})^{2}=\frac{X^{2}}{N^{2}}-\frac{L^{2}}{g_{\phi }}-1,
\label{norm}
\end{equation}
where 
\begin{equation}
X=E-\omega L.
\end{equation}
Equation (\ref{norm}) can be rewritten again to give 
\begin{align}
u^{r}&=\pm \frac{\sqrt{A}}{N}\;Z,  \label{pr} \\
Z^{2}&=X^{2}-N^{2}\Big(\frac{L^{2}}{g_{\phi }}+1\Big) .  \label{z}
\end{align}

For free motion Eqs. (\ref{norm}) and (\ref{pr}) can be obtained, as usual,
as the first integrals of the geodesic equations. If the motion is not
geodesic, the equations remain valid, but $E$ and $L$ are not, in general,
integrals of motion anymore, and should be treated as useful notation only.

Let us denote the components of acceleration in the OO frame by $a_{o}^{(\mu
)}$. For simplicity, we assume hereafter that $A=N^{2}$ (if this is not so,
one can redefine the radial coordinate for motion in the equatorial plane to
achieve this). Then, using Eqs. (112), (116) and (117) of \cite{rad}, we
have 
\begin{align}
a_{o}^{(\phi )}& =-\frac{Z}{\sqrt{g_{\phi }}}L^{\prime };  \label{EQao-phi}
\\
a_{o}^{(t)}& =-\frac{Z}{N}(X^{\prime }+L\omega ^{\prime }) =-\frac{%
Z(E^{\prime }-\omega L^{\prime })}{N};  \label{EQao-t} \\
a_{o}^{(r)}& =-\frac{X}{Z}a_{o}^{(t)}-N\frac{LL^{\prime }}{g_{\phi }}.
\label{EQao-r}
\end{align}

By definition, the energy in the center of mass frame of colliding particles
having unit masses is equal to 
\begin{equation}
E_{c.m.}^{2}=-(u_{1}^{\mu }+u_{2}^{\mu })(u_{1\mu }+u_{2\mu })
=2+2\gamma_{c.m.},  \label{cm}
\end{equation}
where 
\begin{equation}
\gamma _{c.m.}=-u_{1\mu }u_{2}^{\mu }
\end{equation}
is the relative Lorentz factor. Assuming that both particles move towards
the black hole, so that in (\ref{pr}) we should take the minus sign, the
direct calculation, using (\ref{EL}) and (\ref{norm}), gives 
\begin{equation}
\gamma _{c.m.} =\frac{X_{1}X_{2}-Z_{1}Z_{2}}{N^{2}} -\frac{L_{1}L_{2}}{%
g_{\phi}}.  \label{ga}
\end{equation}

\section{Critical and near-critical particles}

In the context of the BSW effect a particle is called ``usual'' if $%
X_{H}\neq 0$ and ``critical'' if $X_{H}=0$ (subscript ``H" here denotes
quantities calculated on the horizon). The effect near extremal horizons is
realized in collision of one usual and one critical particles. Let us
suppose now that there are two particles colliding near a nonextremal
horizon, where 
\begin{equation}
N^{2}\sim \xi,\qquad \xi\equiv \frac{r-r_H}{r_H} .  \label{nx}
\end{equation}

For the critical particle in the horizon limit, according to (\ref{z}), we
would have $Z^2 <0$, which means that it cannot actually reach the horizon 
\cite{prd}: exactly critical particles do not exist in this case. However,
let us then consider a usual particle, for which the expansion 
\begin{equation}
X=X_{H}+x_{1}\xi +x_{2}\xi ^{2}+\cdots
\end{equation}
starts from the first nonvanishing term $X_H \neq 0$. Such a particle
evidently can reach the horizon, and we can choose its point of collision
with another particle close to the horizon $\xi_c \ll 1$ (subscript ``c''
denotes the point of collision). Additionally, we can choose $X_{H}$ to be
small to the same order 
\begin{equation}
X_{H}\sim N_c\sim \sqrt{\xi _{c}}.  \label{xn}
\end{equation}
Such a particle is called ``near critical'', and 
\begin{equation}
Z_H\sim N_c \sim \sqrt{\xi _{c}}.  \label{zn}
\end{equation}
Then, in case the other particle is usual, in accordance with \cite{gp,prd}, their relative Lorentz factor at the point of collision is 
\begin{equation*}
\gamma_{c.m.}\sim N^{-1}_c ,
\end{equation*}
which can be made arbitrarily large by choosing the point of collision $%
\xi_c $ sufficiently close to the horizon and tuning $X_H$ accordingly.

\section{Behavior of the force}

As seen in the previous section, the formal description of the BSW effect on
the kinematic level does not change with the introduction of force \cite%
{gp,prd}. However, in order to understand whether the effect is actually
preserved, we should check i) if it is compatible with the force acting on
particles being \emph{finite} (an obvious physical requirement) and ii)
whether it is possible, given some reasonably arbitrary finite force, to
find/tune the near-critical particle, which is needed for the effect. Here
we will take advantage of the results of analysis performed in \cite{rad}
for extremal horizons to show that the main conclusions remain valid for
nonextremal ones.

The force acting on a particle, which must be bounded, is the one calculated
in that particle's frame. We are interested only in usual particles here, as
critical ones do not exist near nonextremal horizons (see above). The FO
frame \cite{rad} is constructed so that a usual particle's Lorentz factor in
this frame is finite, thus the force acting on the particle in the FO frame
must be finite as well. The OO frame is related to the FO frame through the
Lorentz boost, which is singular in the horizon limit, with $\gamma \sim 1/N
\to \infty$, so the components of the force in the OO frame can diverge.

Let us denote the tetrad components of a particle's acceleration in the FO
frame, which must be finite, as $a_{f}^{(i)}$. The components of
acceleration in the OO frame $a_{o}^{(i)}$ are related with them via the
singular Lorentz boost, and the explicit relations between their asymptotics
are given by Eqs. (68)--(71) of \cite{rad}: 
\begin{align}
a_{f}^{(t)}& =(a_{f}^{(t)})_{0}+(a_{f}^{(t)})_{1}N+O(N^{2});  \label{af-t} \\
a_{f}^{(r)}& =(a_{f}^{(r)})_{0}+(a_{f}^{(r)})_{1}N+O(N^{2});  \label{af-r} \\
a_{o}^{(t)}& =+\frac{(a_{f}^{(t)})_{0}-(a_{f}^{(r)})_{0}}{N}+\big[%
(a_{f}^{(t)})_{1}-(a_{f}^{(r)})_{1}\big]+O(N);  \label{ao-t} \\
a_{o}^{(r)}& =-\frac{(a_{f}^{(t)})_{0}-(a_{f}^{(r)})_{0}}{N}-\big[%
(a_{f}^{(t)})_{1}-(a_{f}^{(r)})_{1}\big]+O(N).  \label{ao-r}
\end{align}
Thus $a_{o}^{(t)}$ and $a_{o}^{(r)}$ can diverge as $1/N$. The $\phi $ and $%
z $ components are the same in the two frames and must be bounded, so
according to Eqs. (72) and (73) of \cite{rad}, 
\begin{equation}
a_{f}^{(\phi )}=a_{o}^{(\phi )}=O(1).  \label{o1}
\end{equation}
This behavior is insensitive to the type of the horizon, extremal or not.

Let us see if this asymptotic behavior is compatible with ``equations of
motion'' (\ref{EQao-phi})--(\ref{EQao-r}) for near-critical particles. Using
the asymptotes (\ref{xn}) and (\ref{zn}) and assuming that $L$, $E$, $L^{\prime}
$ and $E^{\prime}$ are finite, we get 
\begin{align}
& a_o^{(\phi)}\sim \sqrt{\xi};  \label{phi} \\
& a_o^{(t)},a_o^{(r)}\sim 1.  \label{atr}
\end{align}
Here we have omitted the subscript ``c''. We see that the kinematic
restrictions (\ref{ao-t})--(\ref{o1}) are satisfied, and the dynamic
constraints (\ref{phi}) and (\ref{atr}) are even stronger than the kinematic
ones. This means that it is the eqs. (\ref{phi}), (\ref{atr}) that
constitute the actual constraints on the behavior of the force near the
horizon, where collision occurs, for near-critical particles to exist.

In the same way, one can check that for usual particles with $X_H , Z_H \neq
0 $ the dynamic constraints, which follow from (\ref{EQao-phi}) and (\ref{EQao-r}), coincide with the kinematic ones (\ref{ao-t})--(\ref{o1}): 
\begin{equation}
a_o^{(\phi)}\sim 1,\quad a_o^{(r)},a_o^{(t)}\sim 1/N .
\end{equation}

\section{Example: Reissner-Nordstr\"{o}m metric}

For the purely radial motion in the Reissner-Nordstr\"{o}m metric the
equations of motion of a particle with mass $m$ and charge $q$ read 
\begin{align}
ma_{o}^{(t)} =-&\frac{qQ}{r^{2}}\,\frac{Z}{mN}, \\
ma_{o}^{(r)} =+&\frac{qQ}{r^{2}}\,\frac{X}{mN}, \\
& m^{2}a^{2}=\Big(\frac{qQ}{r^{2}}\Big)^{2}.
\end{align}

For near-critical particles with $X\sim Z\sim N$ [(\ref{xn}) and (\ref{zn})], we
get 
\begin{equation*}
a_o^{(t)},a_o^{(r)}\sim 1 ,
\end{equation*}
analogously to (\ref{atr}). For usual particles, $X_H , Z_{H}\neq 0$, so we
have 
\begin{equation*}
a_o^{(t)},a_o^{(r)}\sim 1/N ,
\end{equation*}
which is still allowed, according to Eqs. (\ref{ao-t}), (\ref{ao-r}).

\section{Near-critical particles and effect of dissipation}

The second question that may not be quite clear \textit{a priori} is whether it is
always possible to fine-tune a near-critical particle. Let us suppose that
dissipation is neglected. Then, the solution $x^{\mu }(n)$ is specified by
initial data and for each set of data there exists a single solution.
Instead of fixing conditions at the initial moment of time, however, we can
fix them at the moment when the near-critical particle reaches the horizon: $%
r(0)=r_{H}$, $\dot{r}(0)=X_{H}$ [see (\ref{pr})].

As, by assumption, dissipation is neglected, the system is time symmetric
with respect to time inversion $t\mapsto -t$. Therefore, by integrating
equations of motion back in time, we can recover the trajectory that leads
to near-horizon collision with the unbound $E_{c.m.}$ This is achieved by
taking arbitrarily small $X_{H}$ from the very beginning. Thus the BSW
effect survives in spite of the presence of the force.

In practice, however, a particle can experience the influence of dissipative
forces, such as gravitational radiation reaction. Either dissipation arises
due to terms proportional to velocity (or its higher odd powers) or it
cannot be described in terms of forces at all. What is important,
dissipation violates the symmetry between the two directions of time, which
devaluates the above reasoning. However, if dissipation is small enough, the
presented arguments retain their validity.

Dissipation is small if the time of relaxation $\tau_{diss}$ is much greater
than the characteristic dynamic time scales. In the context of gravity, we
should compare proper time intervals. Let us consider motion of the
near-critical particle. Such a particle moves between the horizon and the
turning point $r=r_{0}$, so collision occurs somewhere within this interval,
which shrinks to zero when $X_{H}\rightarrow 0$ (see for details \cite{gp42} and \cite{circ}).

The proper time of movement until collision is less than the time of
movement from the horizon to the turning point $r=r_{0}$, so its upper
estimate is 
\begin{equation}
\tau _{dyn}=\int\limits_{r_{H}}^{r_{0}}\frac{dr}{|u^{r}|}.  \label{dist}
\end{equation}%
Then, for small $r-r_{H}$, we have 
\begin{equation}
N^{2}\approx 2\kappa (r-r_{H})=2\kappa r_{H}\xi , \qquad 0 \leq \xi_c \leq
\xi \leq \xi _{0},
\end{equation}%
where $\xi _{0}$ corresponds to the turning point. In accordance with (\ref%
{xn}), we have $X_{H}^{2}=\xi _{0}b$, with $b=2\kappa r_{H}(\frac{L^{2}}{%
g_{\phi }}+1)=O(1)$.

Now, it follows from Eqs. (\ref{pr}) and (\ref{z}) that 
\begin{equation}
\tau \approx 2\frac{r_{H}}{\sqrt{b}}\sqrt{\xi _{0}}.
\end{equation}%
Thus the effect of dissipation is small as long as 
\begin{equation}
\frac{\tau _{diss}}{r_{H}}\gg \xi _{c},  \label{dis}
\end{equation}%
which holds automatically for sufficiently small $\xi _{c}$. If a particle
is usual, the effect of dissipation is irrelevant in the context of the BSW
effect at all since it simply transforms a usual trajectory into another
usual one. Thus for near-horizon collision the dissipation effects can be
neglected and cannot restrict the BSW effect, so that the energy in the
center of mass frame can be made arbitrarily large.

It is worth noting that the above discussion does not apply directly to the
case of extremal horizon, since the proper time of reaching the extremal
horizon for a critical particle is infinite. In that case, however, the
existence of the BSW effect is confirmed via different reasoning, based on
the direct analysis of near-horizon trajectories \cite{rad}.

\section{Explicit procedure of tuning}

In this section we demonstrate explicitly, how the procedure of tuning can
be realized for near-critical particles. As the particle is not exactly
critical, tuning should be understood in the approximate sense (small but
nonzero $X_{H}$ on the horizon). As an example, we consider the case of the
azimuthal force, when $a_{o}^{(r)}=0$. Then, it follows from (\ref{EQao-phi}%
) that 
\begin{equation}
g_{\phi }X(X^{\prime }+L\omega ^{\prime })=N^{2}LL^{\prime }  \label{xg}
\end{equation}%
which corresponds to Eq. (134) of \cite{rad}.

Now, we fix small $X_{H}$ and seek the solution in the form of series%
\begin{align}
N^{2}&=2\kappa \xi +\nu _{2}\xi ^{2}+\nu _{3}\xi ^{3}+\cdots \\
\omega &=\omega _{H}-\omega _{1}\xi +\omega _{2}\xi ^{2}+\cdots , \\
g_{\phi }&=g_{H}+g_{1}\xi +g_{2}\xi ^{2}+\cdots , \\
X&=X_{H}+x_{1}\xi +x_{2}\xi ^{2}+\cdots ,  \label{xe} \\
L&=l_{H}+l_{1}\xi +l_{2}\xi ^{2}+\cdots .
\end{align}
Equating the terms of the zeroth order by $\xi $, we obtain from (\ref{xg})
that%
\begin{equation}
x_{1}=\omega _{1}l_{H}.
\end{equation}

The terms of the first order entail 
\begin{equation}
l_{1}=\frac{2g_{H}X_{H}(x_{2}+\omega _{2}l_{H})}{2\kappa
l_{H}+g_{H}X_{H}\omega _{1}}.
\end{equation}%
Repeating the procedure iteratively, we get $l_{2}=l_{2}(X_{H},x_{1},x_{2})$%
, etc. Substituting (\ref{xe}) into (\ref{z}), we find%
\begin{align}
Z^{2}&=X_{H}^{2}-z_{1}\xi , \\
z_{1}&=2\kappa \Big(\frac{l_{H}^{2}}{g_{H}}+1\Big)
\end{align}
where we neglected the term of the order $X_{H}$ in $z_{1}$.

Then, it follows from (\ref{ga}) that%
\begin{equation}
\gamma _{c.m.}\sim \frac{X_{H}-\sqrt{X_{H}^{2}-z_{1}\xi }}{\xi }\text{.}
\end{equation}%
In the region between the horizon and the turning point $0\leq \xi \leq \xi
_{0}=X_{H}^{2}/ z_{1} $ this factor has the order $X_{H}^{-1}$ and can be
made as large as one likes.

\section{Summary}

We have shown that the BSW effect near nonextremal horizons retains its
validity even if the particle experiences the action of forces, provided
some rather weak and reasonable restrictions are imposed on these forces.
For the near-critical particle that plays the crucial role in the effect,
the corresponding conditions are described by Eqs. (\ref{phi}) and (\ref{atr}). In combination with the previous similar results for extremal horizons 
\cite{rad}, this means that the BSW effect turns out to be rather viable and
shows properties of universality. In application of the obtained results
to the issue of gravitation self-force, this should be considered
as the model approach and first approximation only, so the full analysis
requires further study.

\end{document}